\documentclass[fleqn,usenatbib]{mnras}

\usepackage[T1]{fontenc}

\DeclareRobustCommand{\VAN}[3]{#2}
\let\VANthebibliography\thebibliography
\def\thebibliography{\DeclareRobustCommand{\VAN}[3]{##3}\VANthebibliography}

\usepackage{graphicx}	
\usepackage{amsmath}	
\usepackage{amssymb}	
\usepackage{multicol}
\usepackage{wrapfig}
\usepackage{subfig}

\providecommand{\abs}[1]{\lvert#1\rvert}

\usepackage[dvipsnames]{xcolor}

\usepackage{xcolor,colortbl}
\definecolor{Gray}{gray}{0.85}

\newcolumntype{a}{>{\columncolor{Gray}}c}

\usepackage{newtxtext,newtxmath}

\title[eccentricity influences the pebble isolation mass]{How the planetary eccentricity influences the pebble isolation mass}

\author[Chametla et al.]{
Raúl O. Chametla,$^{1,2}$\thanks{E-mail: raul@icf.unam.mx (ROC)}
Frédéric S. Masset,$^{2,3}$
Clément Baruteau$^{4}$ and 
Bertram Bitsch$^{5}$
\\
$^{1}$Charles University, Faculty of Mathematics and Physics, Astronomical Institute, V Hole$\check{s}$ovi$\check{c}$k\'ach 747/2, 180 00 Prague 8, Czech Republic\\
$^{2}$Instituto de Ciencias F\'isicas, Universidad Nacional Autonoma de M\'exico, Av. Universidad s/n, 62210 Cuernavaca, Mor., Mexico\\
$^{3}$University Nice-Sophia Antipolis, CNRS, Observatoire de la C\^ote d'Azur, Laboratoire LAGRANGE, CS 34229. 06304 Nice Cedex 4, France\\
$^{4}$IRAP, Universit\'e de Toulouse, CNRS, UPS, F-31400 Toulouse, France\\
$^{5}$Max-Planck-Institut für Astronomie, Königstuhl 17, 69117 Heidelberg, Germany\\
}

\date{Accepted XXX. Received YYY; in original form ZZZ}

\pubyear{2021}

\begin{document}
\label{firstpage}
\pagerange{\pageref{firstpage}--\pageref{lastpage}}
\maketitle

\begin{abstract}  
We investigate the pebble isolation mass for a planet on a fixed eccentric orbit in its protoplanetary disc by conducting a set of 2D hydrodynamical simulations including dust turbulent diffusion. A range of planet eccentricities up to $e=0.2$ is adopted. Our simulations also cover a range of $\alpha-$turbulent viscosities, and for each pair $\{\alpha,e\}$ the pebble isolation mass is estimated as the minimum planet mass in our simulations such that solids with a Stokes number $\gtrsim 0.05$ do not flow across the planet orbit and remain trapped around a pressure bump outside the planet gap. For $\alpha<10^{-3}$, we find that eccentric planets reach a well-defined pebble isolation mass, which can be smaller than for planets on circular orbits when the eccentricity remains smaller than the disc's aspect ratio. We provide a fitting formula for how the pebble isolation mass depends on planet eccentricity. However, for $\alpha > 10^{-3}$, eccentric planets cannot fully stall the pebbles flow, and thus do not reach a well-defined pebble isolation mass. Our results suggest that the maximum mass reached by rocky cores should exhibit a dichotomy depending on the disc turbulent viscosity. While being limited to ${\cal O}(10\,M_\oplus)$ in low-viscosity discs, this maximum mass could reach much larger values in discs with a high turbulent viscosity in the planet vicinity. Our results further highlight that pebble filtering by growing planets might not be as effective as previously thought, especially in high-viscosity discs, with important implications to protoplanetary discs observations.
\end{abstract}




\begin{table*}
\caption{Initial conditions and main parameters of our simulations ($a_{\rm p}$ denotes the planet's fixed semi-major axis).}
    \label{tab:parameters}
\begin{tabular}{ p{8.5cm}||p{8.5cm}  }
 \hline
 GAS & DUST\\
 \hline
 Aspect ratio \hspace{2.9
 cm}        $h$  \hfill 0.04& Dust size \hspace{3.3cm} $s\mathrm{[mm]}$  \hfill  $[1-300]$\\
 $\alpha$-turbulent viscosity \hspace{1.85cm} $\alpha$  \hfill  $[10^{-4}-10^{-2}]$& Internal density of particles \hspace{1.25cm} $\mathit{\rho_{\rm p}\mathrm{[g/cm^3]}}$  \hfill  2 \\
 Mass of the central star \hspace{1.6cm} $M_{\star}[M_{\odot}]$  \hfill  1.0&  Number of dust particles \hspace{1.5cm} $N_{\rm p}$  \hfill  50 000  \\
Number of cells in radius \hspace{1.35cm}           $N_r$  \hfill    384& Size distribution \hspace{2.4cm} $n(s)$  \hfill  $\propto s^{-1}$ \\
Number of cells in azimuth \hspace{1.15cm}           $N_{\varphi}$  \hfill    1206& Stokes number \hspace{2.6cm} $St$  \hfill  $[3\times10^{-4}-0.2]$   \\
 Radial extent of the grid \hspace{1.5cm}           $r[a_{\rm p}]$  \hfill    [0.5-2.5]& Dust turbulent diffusivity  \hspace{1.45cm}           $D_d$  \hfill    $\nu\dfrac{1+4St^2}{(1+St^2)^2}$ \\
Azimuthal extent of the grid \hspace{1.05cm}           $\varphi$  \hfill    [$0-2\pi]$&  \\
 Surface density at $a_{\rm p}$ \hspace{1.9cm}           $\Sigma[\mathrm{g/cm^2}]$  \hfill    900&   \\
 Surface density slope \hspace{1.85cm}           $\sigma$  \hfill    -1/2&   \\
 Temperature slope \hspace{2.2cm}           $\beta$  \hfill    -1&   \\
 \hline
\end{tabular}
\end{table*}

\section{Introduction}
Planets beyond a few Earth masses can generate a partial gap around their orbit in the disc of gas and dust in which they form. Depending on the disc's temperature and turbulent viscosity, a pressure bump can build up at the outer edge of this gap \citep[e.g.,][]{Crida_etal2006,2006A&A...459L..17P,2015ApJ...807L..11D}. Several studies have shown that for a planet on a circular orbit, this pressure bump can act as a barrier against the radial drift of solids marginally coupled to the disc gas, thereby halting their accretion on the planet \citep{2012A&A...546A..18M,Lambrechts_etal2014,2014A&A...572A.107L,2017AJ....153..222B,Ataiee_etal2018,Bitsch_etal2018}. Such solids have a size comparable to that of pebbles, and the critical mass that a planet needs to reach before it gets completely isolated from surrounding pebbles in the disc is called the pebble isolation mass (PIM).

Since the PIM depends on the effective trapping of pebbles around a pressure bump outside the planet orbit, previous studies have examined its dependence on the aspect ratio $h$ and the $\alpha$ turbulent viscosity of the disc gas, as well as on the Stokes number ($St$) of the solids. The effect of the planet eccentricity ($e$) on the PIM has not been previously addressed, which is yet of importance for planet formation and evolution models \citep[][]{PMID:26289203,2019A&A...627A..83L,2019A&A...623A..88B,2020A&A...643A..66B,2021A&A...650A.116M,Izidoro21}. It also has a strong observational motivation. While the mass distribution of planets in observed close-in planetary systems could be accounted for by the disc's thermal mass \citep[][]{2019ApJ...874...91W} or the PIM \citep[][]{2019A&A...630A..51B}, most of these planets actually have slightly eccentric orbits \citep[][]{2011arXiv1109.2497M,2011ApJS..197....8L,2012ApJ...758...39J,2012ApJ...761...92F,2016PNAS..11311431X,2018ApJ...860..101Z} which naturally leads to wonder whether and how the PIM differs for eccentric planets. For a protoplanetary core on an eccentric orbit, \citet{LO2018} found that the pebble accretion efficiency first increases with $e$ once $e$ becomes of the order the absolute value of the dimensionless radial pressure gradient $\abs{\eta} = (h^2/2) \abs{ \partial\log P / \partial\log r }$. However, when $e$ exceeds a few percent typically, that is of order $h$, the high relative velocity between the planet and pebbles precludes pebble accretion, whose efficiency drops quickly. 

Although previous studies have shown that the eccentricity of a planet can be dampened by the interactions with the gas disc \citep[][]{2008A&A...482..677C,2010A&A...523A..30B}, several mechanisms can grow and maintain a substantial level of eccentricity for planets that are still embedded in their protoplanetary disc. Before they reach their PIM, low-mass accreting planets can acquire an eccentricity comparable to the disc's aspect ratio if their luminosity (which results from the heat release due to solids accretion) exceeds a critical value \citep[][]{2017MNRAS.469..206E,Chrenko2017,2019MNRAS.485.5035F,2021MNRAS.tmp.3059R}. Eccentricity growth can also arise from gravitational interactions with another planet companion, and a large range of eccentricities can be obtained depending on the planets mass, their mutual separation, and the disc's physical properties like its density and temperature (see, e.g., \citealp{Kley04} and \citealp{Pierens08} for resonant interactions in different regimes of planet masses). While giant planets on circular orbits have a mass largely beyond the PIM, this might not be the case, as we will see, when such planets are on eccentric orbits. Single massive planets, typically above a Jupiter mass, can experience eccentricity growth due to the dominance of eccentric resonances in the disc-planet interaction, particularly when the planet is in a low-density environment like a gas cavity \citep[e.g.,][]{2001A&A...366..263P,2006A&A...447..369K,2006ApJ...652.1698D,2013A&A...555A.124B,Debras21}. For instance, the simulation of \citet{Baruteau2021} shows that a 2 Jupiter mass planet around a 2 Solar-mass star can reach and maintain an eccentricity $\sim$0.25 after migrating into a low-density gas cavity in a few Myr (see their figure 1). Moreover, planets eccentricity can be grown and maintained via the interaction between the planet and a distant vortex, as shown in Chametla $\&$ Chrenko (submitted).

With all this in mind, we revisit in this study the pebble isolation mass in a protoplanetary disc by considering planets with eccentricities between 0 and 0.2, using a disc model very similar to that of \citet{Ataiee_etal2018}, who studied the pebble isolation mass in 2D hydrodynamical simulations including solid particles for planets on circular orbits. The planet's eccentricity is fixed in a given simulation, and we do not assume one particular mechanism behind the growth and maintenance of the eccentricity. The paper is organized as follows. In Section \ref{sec:model} we present the physical model and numerical setup used in our 2D gas and dust hydrodynamical simulations. In Section \ref{sec:previously} we present a brief summary of the results of two previous studies on the isolation mass for planets on circular orbits. We present our results in Section \ref{sec:results}, followed by Section~\ref{sec:fit} in which we introduce a fitting formula for the pebble isolation mass of eccentric planets. Concluding remarks can be found in Section \ref{sec:conclusion}.

\section{Physical Model and numerical setup}
\label{sec:model}
We have carried out two-dimensional (2D) hydrodynamical simulations to model the evolution of the gas and dust in protoplanetary discs harboring planets on eccentric orbits. Apart from the planet eccentricity, the physical model and numerical setup used in this study basically follow those in \citet{Ataiee_etal2018}, which we briefly recall for convenience. The simulations use the code Dusty FARGO-ADSG, which is an extended version of the original 2D FARGO code \citep[][]{Masset2000} that includes Lagrangian dust particles \citep{BZ2016}.

For the gas, the continuity and Navier-Stokes equations are solved on a polar grid centred on to the star, and in a frame co-rotating with the planet. For simplicity, a locally isothermal equation of state is used with the vertically integrated gas pressure $P$ and the isothermal sound speed $c_{\rm s}$ satisfying $P=\Sigma c_{\rm s}^2$, with $c_{\rm s}$ kept stationary. We also note that the inclusion of an energy equation in \citet{Bitsch_etal2018} does not change significantly the PIM in their simulations. The source terms in the Navier-Stokes equation include the gravitational potential due to the star, the planet and the indirect terms arising from the stellar acceleration imparted by the disc gas and the planet. In the planet potential, a softening length $\epsilon=0.4H(r_{\rm p})$ is included to mimic the effects of finite vertical thickness, with $H$ the disc's pressure scale height and $r_{\rm p}$ the planet's orbital radius. A viscous acceleration is also included with a turbulent kinematic viscosity ($\nu$) that is expressed in terms of a dimensionless alpha parameter ($\nu = \alpha c_{\rm s}H)$. The initial radial profiles of the surface density and temperature are power-law functions with exponents $\sigma$ and $\beta$, respectively. 

On the other hand, the dust (or solids) in the disc are modelled as Lagrangian test particles in our simulations. The gas drag acceleration is accounted for, so is dust turbulent diffusion via stochastic kicks on the particles position (see Section 2.1 in \citealp{Ataiee_etal2018} for more details). The dust drag acceleration on the gas is, however, discarded. Table~\ref{tab:parameters} summarizes the set of disc parameters and initial conditions used in our simulations. We recall that the Stokes number of the solid particles is the ratio between the stopping time (i.e., the time particles need to adjust to the gas flow due to drag forces) and the dynamical time. In the Epstein regime, which is relevant to our disc model, it can be expressed as $St=\pi s\rho_{\rm p}/(2 \Sigma)$, where $s$ and $\rho_{\rm p}$ are the size and internal density of the dust particles, respectively.

For the boundary conditions, we use radial damping boundaries to avoid reflections of the planet wakes near the radial edges of the grid in the same way as described in \citet{deVal_etal2006}. The planet mass is gradually increased over the first five orbits of the simulations to avoid artefacts due to the instantaneous insertion of the planet in the disc. Before inserting the dust particles beyond the planet orbit in the disc, we first evolve the gas disc over 1000 to 6000 planet orbits depending on the disc viscosity, to achieve a steady state. We subsequently let the dust evolve in the disc for up to 3 times the duration of the preliminary simulations, which we find necessary to ascertain whether solid particles with $St>0.05$ get effectively trapped or not beyond the planet orbit. The main differences between our setup and that in \citet{Ataiee_etal2018} are in the number of grid cells and the total duration of the simulations.

\section{Summary of previous results on the pebble isolation mass for circular planets}
\label{sec:previously}
\citet{Ataiee_etal2018} estimated the PIM for circular planets via a semi-analytical calculation using the analytical gap density profile of \citet{2015ApJ...807L..11D}. Their PIM, which takes the following expression,
\begin{equation}
\frac{M_{\mathrm{PIM}}}{M_{\star}}\approx h^3\sqrt{37.3\alpha+0.01}\times \left[1+0.2\left(\frac{\sqrt{\alpha}}{h}\sqrt{\frac{1}{St^2}+4}\right)^{0.7}\right],
 \label{eq:Ataiee}
\end{equation}
\noindent was checked against 2D gas and dust hydrodynamical simulations with a setup that is very similar to that used in the present study. A good agreement was found, except at large turbulent viscosities for which the effective PIM found in their simulations was up to a factor $\sim$3 larger than predicted by Eq.~(\ref{eq:Ataiee}) for $\alpha=10^{-2}$ (see Figure 8 in \citealp{Ataiee_etal2018}).

By means of 3D hydrodynamical simulations, \citet{Bitsch_etal2018} found a PIM that could be expressed as (see Eq. A6 in \citealp{Ataiee_etal2018})
\begin{equation}
M_{\mathrm{PIM}} \approx 25f_{\mathrm{fit}}M_{\oplus} \times \left(1 + 4.2\frac{\alpha}{St}\right),
 \label{eq:B+2018_0}
\end{equation}
with
\begin{equation}
f_{\mathrm{fit}}=\left[\frac{h}{0.05}\right]^3\left[0.34\left(\frac{\log(\alpha_3)}{\log(\alpha)}\right)^4+0.66\right]\left[1-\frac{\frac{\partial \ln P}{\partial \ln r}+2.5}{6}\right]
 \label{eq:B+2018_1}
\end{equation}
and $\alpha_3=10^{-3}$. A detailed comparison between Eqs.~(\ref{eq:Ataiee}) and~(\ref{eq:B+2018_0}) can be found in appendix A of \citet{Ataiee_etal2018}.

\section{Results of hydrodynamical simulations}
\label{sec:results}
\begin{figure}
	\includegraphics[scale=0.5]{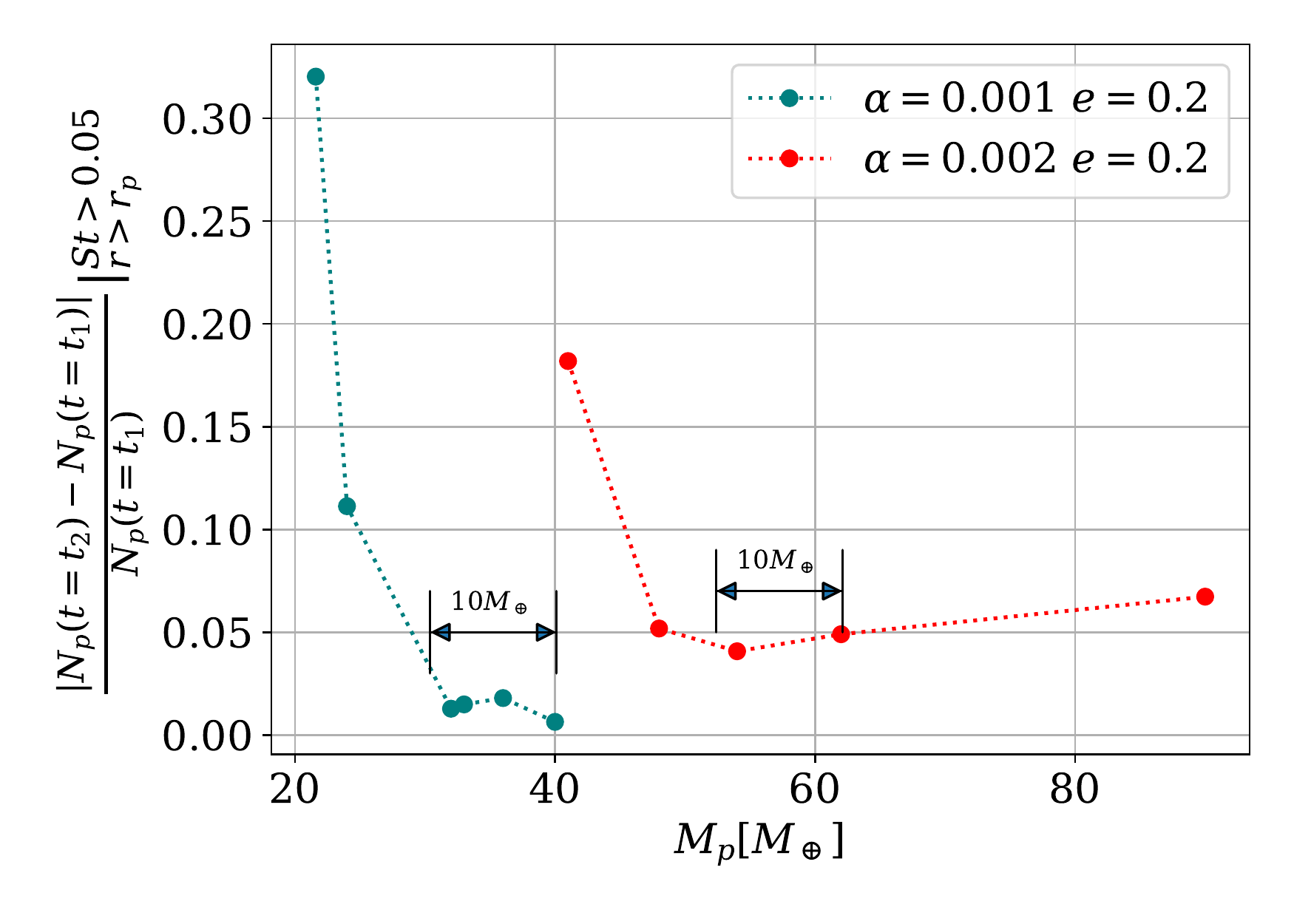}
    \caption{Relative difference between $t_1 = 1000$ and $t_2 = 2000$ orbits of the number of solid particles with $St >0.05$ that stay beyond the planet orbit, as a function of planet mass. Results are obtained for $e=0.2$, and for $\alpha = 10^{-3}$ (green) or $\alpha = 2\times 10^{-3}$ (red).}
    \label{fig:criterion}
\end{figure}

Following \citet{Ataiee_etal2018}, we consider that a planet has reached its PIM when no solids with a Stokes number $\gtrsim 0.05$ can flow through the planet orbit, being effectively trapped around the pressure bump outside the planet gap. As long as $\alpha\leq10^{-3}$, we find that this numerical criterion can be satisfied across the range of eccentricities that we have explored. However, for larger viscosities, we find that there are always solids with $St \gtrsim 0.05$ that can cross the planet orbit when it is eccentric (and only when it is eccentric). For such large viscosities, the flux of $St>0.05$ particles leaving the pressure bump and flowing through the planet orbit, which we now simply refer to as the pebbles flux, first decreases with increasing planet mass (as expected), and then stays approximately the same when further increasing planet mass. This is illustrated in Fig.~\ref{fig:criterion}, which shows the relative difference between 1000 and 2000 planet orbits of the number of solid particles with $St >0.05$ that stay beyond the planet orbit, and which can be seen as a measure of the pebbles flux.

Reaching or not the PIM depends on how large the remaining pebbles flux is or, said differently, how efficient the planet filtration is. Since our simulations do not use a realistic particles size distribution (see Table~\ref{tab:parameters}), defining a quantitative criterion for when a planet reaches its PIM can be a little arbitrary. In the end, we consider that the PIM is reached when the pebbles flux does not increase with increasing planet mass over $10M_\oplus$ from the first planet mass for which the pebbles flux reaches a near minimum. This criterion is highlighted in Fig.~\ref{fig:criterion}.
For $\alpha=0.001$ and $e=0.2$ (green), we see that the pebbles flux decreases from about 27$M_\oplus$ to $40M_\oplus$, where it becomes near zero. In that case, we consider that the planet reaches its PIM and that the latter is $40M_\oplus$. However, for $\alpha=0.002$ and $e=0.2$, the pebbles flux continuously increases from about 54$M_\oplus$ to $90M_\oplus$, so we consider that the planet does not strictly reach its PIM. An approximate PIM can still be inferred, which in that case we consider equal to $90M_\oplus$.

\begin{figure*}
	\includegraphics[scale=0.65]{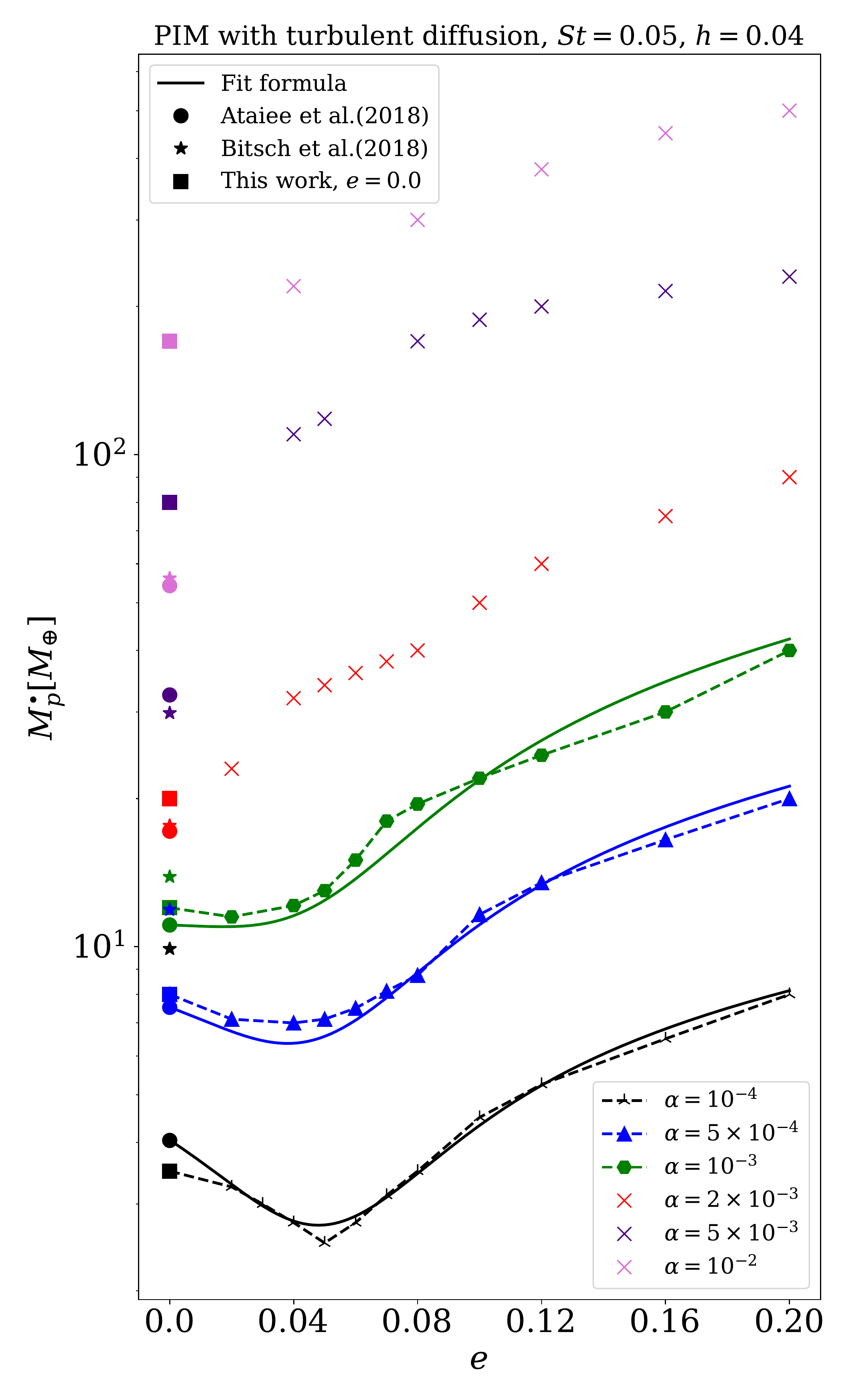}
    \caption{Pebble isolation mass $M_p^{\bullet}$ versus eccentricity $e$ for six values of the $\alpha$ turbulent viscosity. The disc aspect ratio $h$ is fixed to 0.04 and the Stokes number $St$ to 0.05. Only for $\alpha \leq 10^{-3}$ do we find a well-defined PIM for all eccentricities, and we show by solid curves the predictions of our fitting formula given by Eq.~(\ref{eq:IMEP}). Otherwise, we show by X symbols approximate values of the PIM for which there is still pebbles flowing through the planet orbit.
    }
    \label{fig:ecc}
\end{figure*}

Having refined the criterion for when a planet reaches its PIM, we show in Fig.~\ref{fig:ecc} the PIM (hereafter, $M_p^{\bullet}$) inferred from our hydrodynamical simulations for different planet eccentricities. We distinguish the results for $\alpha\leq10^{-3}$, which are shown by dashed curves and for which a well-defined PIM is found, from those for $\alpha\geq2\times10^{-3}$, which are shown by X symbols, and for which we find only an approximate PIM when the planet eccentricity is not zero. 

\begin{figure*}
	\includegraphics[scale=0.6]{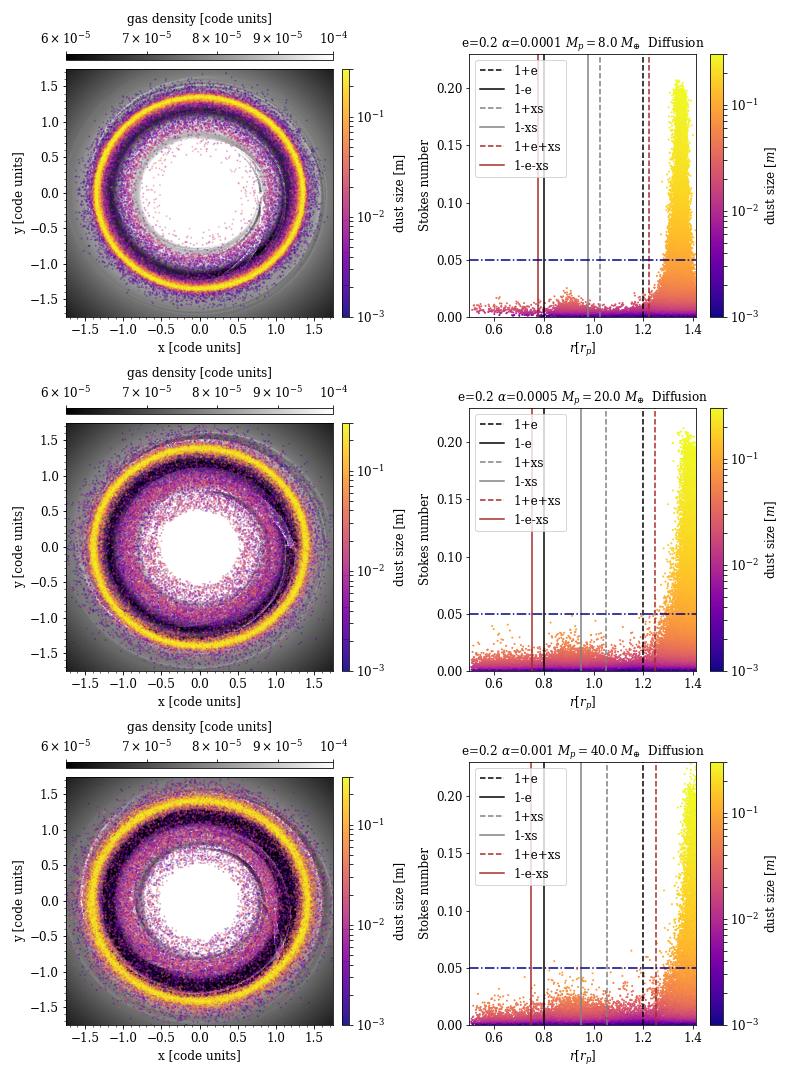}
    \caption{Results of simulations for $e=0.2$ and $\alpha\leq10^{-3}$, when the planet reaches its pebble isolation mass (the planet mass is, from top to bottom, $M_p=8M_{\oplus}$, $M_p=20M_{\oplus}$ and $M_p=40M_{\oplus}$). \textit{Left}: gas surface density with the position of the solid particles overplotted by coloured dots.  \textit{Right}: Stokes number of the solid particles versus orbital radius $r$ and particles size.
    }
    \label{fig:r_vs_St_IMEP}
\end{figure*}
\begin{figure*}
	\includegraphics[scale=0.6]{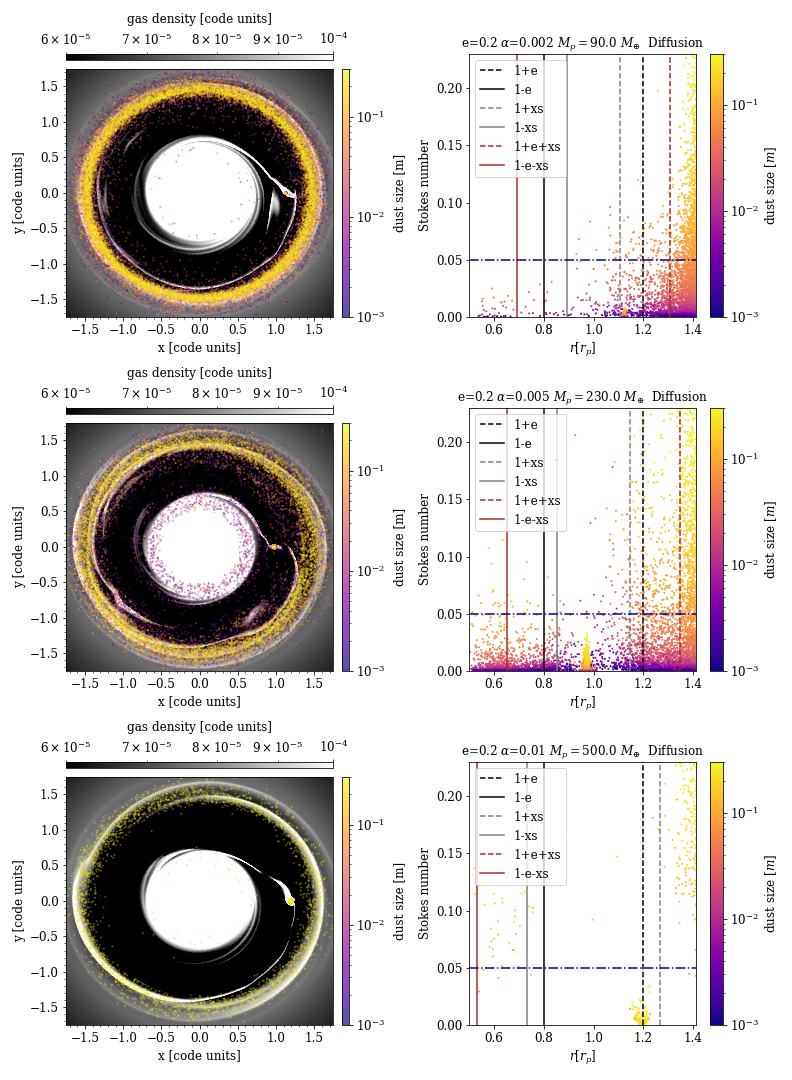}
    \caption{Continuation of Fig.~\ref{fig:r_vs_St_IMEP} for $\alpha>10^{-3}$, i.e., for the cases where the planet does not strictly reach an isolation mass in terms of solids with $St > 0.05$. The corresponding planet mass is $90M_{\oplus}$, $230M_{\oplus}$ and $500M_{\oplus}$ from top to bottom.}
    \label{fig:gas_dust2}
\end{figure*}

Before describing our results for eccentric planets in more details, we stress that for $e=0$ the difference between the PIM inferred from our simulations (square symbols in Fig.~\ref{fig:ecc}) and that predicted by the semi-analytical prediction of \citet{Ataiee_etal2018} (filled circles, see Eq.~\ref{eq:Ataiee}) is very similar to that already reported in Figure~8 of \citet{Ataiee_etal2018}. For $e=0$ we thus recover the same results as \citet{Ataiee_etal2018} despite the few minor differences in the numerical setup of the simulations. Note also the overall good agreement between Eqs.~(\ref{eq:Ataiee}) and~(\ref{eq:B+2018_0}), especially at large viscosities (compare the filled circles and star symbols in Fig.~\ref{fig:ecc}; the quantity $\partial\ln P / \partial\ln r$ in the rightmost square bracket in Eq.~\ref{eq:B+2018_1} was substituted by $\sigma+\beta$ in light of our 2D disc model). Again, the reader is referred to the appendix A of \citet{Ataiee_etal2018} for a detailed comparison between Eqs.~(\ref{eq:Ataiee}) and~(\ref{eq:B+2018_0}).

Now, moving to our results for eccentric planets, for our two lowest $\alpha$-viscosities, we find that the PIM first decreases with $e$, up to $e\sim h$, and it then increases monotonically for the range of eccentricities that we have considered ($e<0.2$). In particular, for $\alpha=10^{-4}$ and $e=0.05$, $M_p^{\bullet} \approx2.5M_{\oplus}$, which is $1M_{\oplus}$ less than the PIM in the circular case. A tentative explanation for this trend is as follows. When the eccentricity first increases, the relative velocity between the planet and the background gas increases, and so does the strength of the shock imparted by the planet's outer wake to the background gas. A moderately eccentric planet can thus provide a similar damping of angular momentum (or of potential vorticity) as a slightly more massive but circular planet. This is why the PIM first decreases upon increasing $e$. However, when the eccentricity is large enough, the planet has a larger excursion through the disc, and can deposit a smaller proportion of angular momentum to the background disc. It is as if the planet's outer wake was smoother or weaker. This explains why the PIM ultimately increases with $e$ at large $e$.

Our results for $\alpha \leq 10^{-3}$ are further illustrated with the help of Fig.~\ref{fig:r_vs_St_IMEP}, which shows the position of the solid particles in the simulations superimposed on the gas surface density, as well as the Stokes number of the particles versus their orbital radius $r$. In the left column of the figure, the formation of a ring of particles for the largest sizes ($\gtrsim 0.1$ m) is clearly seen beyond the orbit of the planet, around the location of the pressure bump built up by the planet's outer wake. Even if the planet has reached its PIM based on our criterion, solids with $St \lesssim 0.05$ can still flow across the planet orbit and drift towards the central star, as can be seen in the right column of Fig.~\ref{fig:r_vs_St_IMEP}. We note that, if the solids with $0.01 < St < 0.05$ passing through the planet orbit were accreted with a 100$\%$ efficiency onto the planet, their contribution to the increase in the planet mass would be very small, as we have verified -- around five orders of magnitude less than for solids with $St>0.05$. It is also in line with the fact that pebble accretion onto planetary cores should be most efficient for solid particles with Stokes numbers in the range $[0.1-1]$ \citep[for a review see, e.g.,][]{2016SSRv..205...77B, 2017AREPS..45..359J}. We therefore think that the threshold Stokes number of 0.05 that we apply for the practical estimation of the PIM via our simulations is an appropriate value. It is also interesting to note the formation of a secondary ring of solid particles with $St<0.05$ slightly inward of the mean orbit of the planet when the latter is eccentric (see discussion below).

The previous trend for how the PIM changes with $e$ does not hold for $\alpha \gtrsim 10^{-3}$, and we now find that the PIM increases monotonically with $e$. We think that there are two good reasons for this. First, increasing turbulent viscosity shortens the viscous diffusion timescale across the wakes, and it therefore smooths out more efficiently the shocks induced by the wakes, even when the planet is moderately eccentric. Second, a larger viscosity goes with a larger dust turbulent diffusivity, so that it is even easier for solid particles to get kicked out from a (weaker) pressure bump. Furthermore, recall that for $\alpha \gtrsim 10^{-3}$, eccentric planets do not reach a well-defined PIM because of the persistence of a flux of $St > 0.05$ particles crossing the planet orbit. This is actually illustrated by Fig.~\ref{fig:gas_dust2}, where it is clear that some particles with $St > 0.05$ can flow across the planet orbit, implying that their accretion by the planet may not be fully stalled. Although an eccentric planet cannot fully stop the flow of $St>0.05$ particles for $\alpha \gtrsim 10^{-3}$, this does not necessarily mean that the planet will accrete a substantial amount of pebbles. In that regard we mention the recent study of \citet{LO2018}, who have shown that dust turbulent diffusion could considerably reduce the efficiency of pebble accretion. An analysis of how an eccentric planet would effectively accrete pebbles in our hydrodynamical simulations is beyond the scope of this work. We note that an accumulation of solid particles in the planet's circumplanetary environment is clearly seen for the three simulations shown in Fig.~\ref{fig:gas_dust2}.

\begin{figure}
   \includegraphics[width=1.0\linewidth]{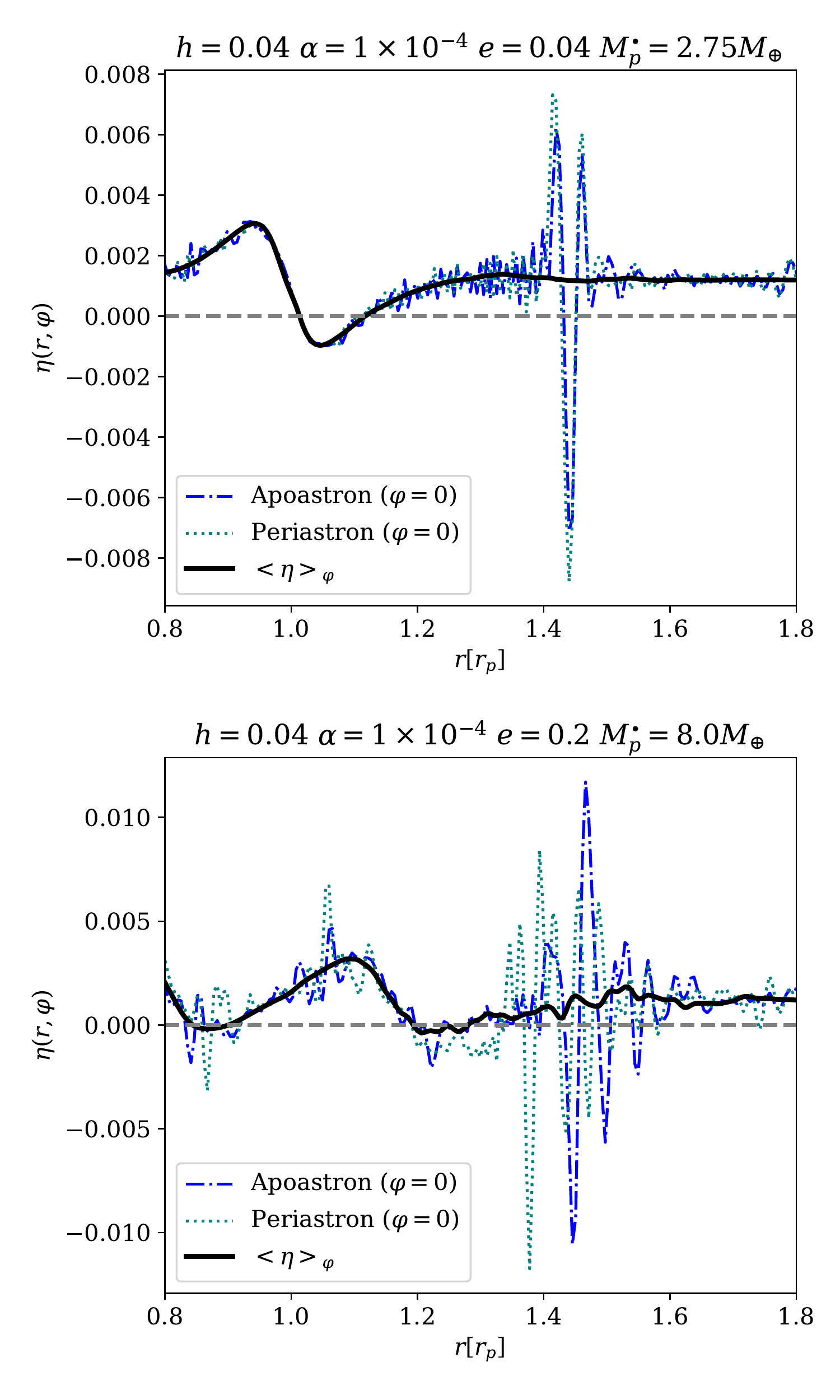}
  \caption{Dimensionless pressure gradient parameter $\eta(r,\varphi)$ (see Eq.~\ref{eq:eta_p}) versus orbital radius $r$ at $\varphi=0$ when the planet is either at its periastron or apoastron. Overplotted as a thick solid curve is the azimuthally-averaged profile of $\eta$ at apoastron. The planet eccentricity is $e=0.04$ (top) and $e=0.2$ (bottom), the disc turbulent viscosity is $\alpha=10^{-4}$ in both panels.}
  \label{fig:eta}
\end{figure}

Before closing this section, we note that the formation of a pressure bump outside the planet gap can be conveniently inferred, from the simulations, by the dimensionless quantity
\begin{equation}
\eta\equiv-\frac{h^2}{2}\frac{\partial\log P}{\partial \log r},
 \label{eq:eta_p}
\end{equation}
which is a way to quantify how strong the radial pressure gradient is. For a circular planet which has just reached its PIM, the $\eta$ radial profile has a smooth behaviour with only a single null (or slightly negative) value outside the planet orbit, as can be seen in Figures 1 and 4 of \citet{Ataiee_etal2018} and \citet{Bitsch_etal2018}, respectively. For an eccentric planet that reaches its PIM, however, we find that the $\eta$ radial profile features several radii with null (or slightly negative) values near the planet orbit as can be seen in Fig.~\ref{fig:eta}. However, we find that only the orbital radii where the azimuthally-averaged $\eta$ profile becomes null or negative slightly beyond the planet orbit can form robust pressure bumps that are able to stall the pebbles flux, as can be seen in Fig.~\ref{fig:r_vs_St_IMEP}. In other words, although $\eta$ may well depend on azimuth, and have null or negative values at different azimuthal angles and different orbital radii (see dotted and dash-dotted curves in Fig.~\ref{fig:eta}), these locations do not constitute solid pressure bumps that are able stop the pebbles flux.

\begin{figure}
   \includegraphics[width=1.0\linewidth]{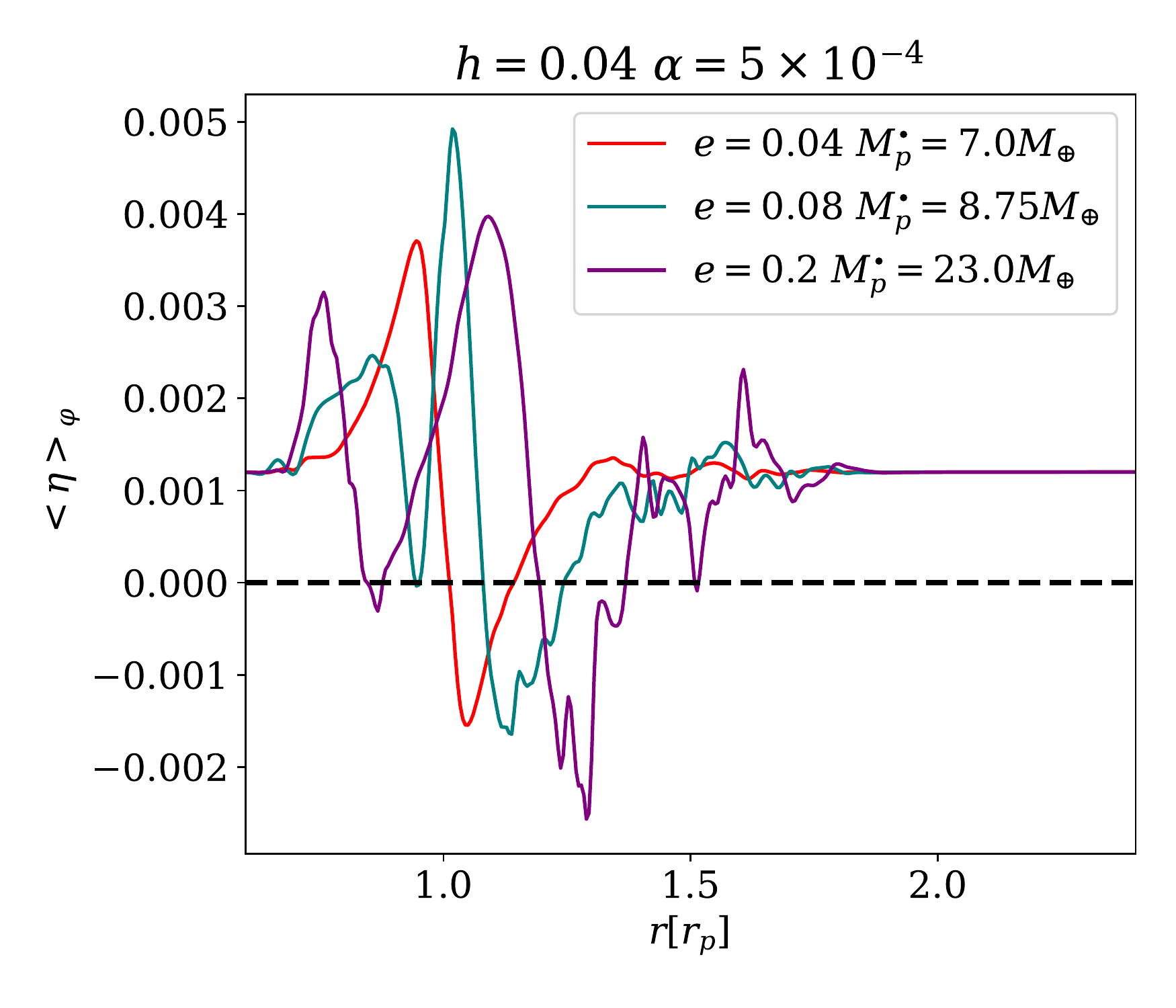}
  \caption{Azimuthally-averaged radial profile of the dimensionless pressure gradient $\eta$ (see Eq.~\ref{eq:eta_p}) when the planet is at apoastron, for $\alpha=5\times10^{-4}$ and three different eccentricities. The radial locations where this quantity cancels out correspond to pressure maxima that may trap solid particles, just like for circular planets, except that eccentric planets can feature several such locations.}
  \label{fig:eta2}
\end{figure}

\begin{figure}
   \includegraphics[width=1.1\linewidth]{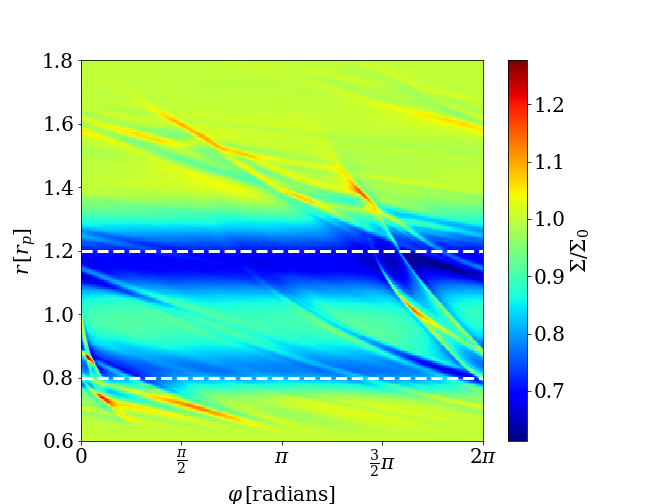}
  \caption{Distribution of the gas surface density $\Sigma(r,\varphi)$ normalized by $\Sigma_0$ in polar coordinates for a planet with mass $M_p^{\bullet}=8.0M_\oplus$ on an $e=0.2$ eccentric orbit in an $\alpha=10^{-4}$ disc. White dotted lines represent the radial positions at $r_{\rm p}(1\pm e)$, respectively.}
  \label{fig:two_gaps}
\end{figure}

Furthermore, we show in Fig.~\ref{fig:eta2} the azimuthally-averaged radial profile of $\eta$ (hereafter, $\langle\eta\rangle_{\varphi}$) when an eccentric, PIM-reaching planet is at apoastron for $\alpha=5\times10^{-4}$ and three values of the planet eccentricity. It can be clearly seen that when the eccentricity increases, there can be several radii at which $\langle\eta\rangle_{\varphi}$ cancels out, and which thus constitute pressure traps where the flow of solid particles can be stalled. In particular, for $e=0.2$, a pressure bump forms inward of the planet orbit. It arises from the fact that, when an eccentric planet carves a gap there is a range of azimuthal angles where gas can pass through the gap and be accumulated over the radial extent $r_{\rm p}(1 \pm e)$ (see Fig. \ref{fig:two_gaps}). The planet actually opens a secondary gap near $r=r_{\rm p}(1-e)$, which is a little shallower than the main gap.

\section{A simple fitting formula for the pebble isolation mass for eccentric planets}
\label{sec:fit}
We have shown in the previous section that only for $\alpha\leq10^{-3}$ does an eccentric planet (strictly) reach a PIM. With this in mind, we provide in this section a fitting expression for the dependence of the PIM with the eccentricity $e$ of the planet orbit. As shown by the solid curves in Fig.~\ref{fig:ecc}, our simulations results are well reproduced by the following expression:
\begin{equation}
M_p^{\bullet}(e) = M_p^{\bullet}(e=0) \times \left\{1 + (400\alpha+0.2)\frac{e}{h} - \frac{5e/h }{\left[8+\left(\dfrac{e}{h}\right)^3\right]}\right\},
\label{eq:IMEP}
\end{equation}
\noindent with $M_p^{\bullet}(e=0)$ given by Eq.~(\ref{eq:Ataiee}). The relative difference between the PIM inferred from our simulations and that given by Eq.~(\ref{eq:IMEP}) does not exceed $10\%$.

We point out that our new expression for the PIM includes a dependence with the disc aspect ratio $h$ and the Stokes number $St$, while our simulations were restricted to $h=0.04$ and computed the isolation mass for a threshold Stokes number of 0.05. It can thus be seen as an extension of the results of the semi-analytical derivation of \citet{Ataiee_etal2018} to non-zero eccentricities. This approach is justified by the good agreement obtained by \citet{Ataiee_etal2018} between the prediction of their semi-analytical derivation and their results of hydrodynamical simulations, as already mentioned above (see their Figure~8).

\section{Discussion and Conclusions}
\label{sec:conclusion}
In this work we report the results of a set of 2D gas and dust hydrodynamical simulations including dust turbulent diffusion with the aim to determine the pebble isolation mass for planets on eccentric orbits in their protoplanetary disc. As stated in the introduction, several mechanisms may cause a planet to reach a substantial level of eccentricity in its disc. Our simulations do not focus on one such mechanism. Instead, a range of planet eccentricities is adopted that are kept constant in each simulation. It is however interesting to consider the case recently studied \citep[][]{2017MNRAS.469..206E,2019MNRAS.485.5035F} of low-mass planets made eccentric by the disturbance created in the disc by their own accretion luminosity. In such case, accretion may proceed in an intermittent manner in a low-viscosity disc: eccentric planets that reach a lower PIM stop accreting, their luminosity decays, and so does their eccentricity, so that pebbles start flowing inwards again. Our study generalizes the previous works by \citet{Bitsch_etal2018} and \citet{Ataiee_etal2018} which were for planets on fixed circular orbits.

For $\alpha$-turbulent viscosities less than $10^{-3}$, our simulations show that eccentric planets reach a well-defined pebble isolation mass, at least for the range of eccentricities that we have probed ($e\leq0.2$). How the pebble isolation mass depends on eccentricity is also related to the turbulent viscosity. When the latter is small enough, the pebble isolation mass first decreases with $e$ so long as $e \lesssim h$ and increases beyond. As $\alpha$ approaches $10^{-3}$ the isolation mass increases rather smoothly with $e$. For $\alpha > 10^{-3}$, however, eccentric planets never fully contain the flow of pebble particles beyond the planet orbit, and such planets do not strictly reach an isolation mass. This is in contrast to circular planets, for which a well-defined pebble isolation mass is found even at large turbulent viscosities. This may have interesting implications for the formation of the cores of massive planets in highly turbulent protoplanetary discs, with the possibility of a reduced efficiency of accretion for eccentric planets \citep[][]{LO2018,2021A&A...647A.175O}. It would be interesting to examine whether such planets could still accrete gas with their envelope contraction not being hindered by the accretion of solids. 

Furthermore, our results imply that the masses of potential planets derived via hydrodynamical simulations designed to match ALMA observations (e.g. \citealt{Zhang2018}) might be off by a factor of $\sim$2, if the planets harbor low eccentricities (see also \citealp{YXChen21}). In addition, our simulations show that pebble filtering by slightly eccentric planets embedded in discs with large turbulent viscosity is not efficient, implying that either Jupiter retained a circular orbit or that the Solar System's disc harbored a low viscosity to allow Jupiter's core to separate the reservoirs of carbonaceous and non-carbonaceous chondrites \citep{Kruijer2017}.

Lastly, we provide a fitting expression for the dependency of pebble isolation mass with eccentricity which could be helpful for models of population synthesis considering cores of massive planets on orbits with low-to-moderate eccentricities \citep[][]{PMID:26289203,2016ApJ...825...63C,2019A&A...623A..88B,2020A&A...643A..66B,2021A&A...650A.116M,Izidoro21}.

\section*{Acknowledgements}

We thank our referee, Elena Lega, for a constructive report. The work of R.O.C. was supported by the Czech Science Foundation (grant 21-23067M) and also acknowledges a postdoctoral CONACyT grant with number 769351. F.M. gratefully acknowledges support from grants UNAM-DGAPA-PASPA and UNAM-DGAPA-PAPIIT IG-101-620, as well as the University of Nice-Sophia Antipolis and the Observatoire de la C\^ote d'Azur for hospitality. Computational resources were available thanks to a Marcos Moshinsky Chair. B.B. thanks the European Research Council (ERC Starting Grant 757448-PAMDORA) for their financial support.

\section*{Data Availability}

The FARGO-ADSG code is available from \href{http://fargo.in2p3.fr/-FARGO-ADSG-}{http://fargo.in2p3.fr/-FARGO-ADSG-}. The code version including a Lagrangian treatment of the dust particles, Dusty FARGO-ADSG, can be made available for use on a collaborative basis upon request to C. Baruteau. The input files for generating our hydrodynamical simulations will be shared on reasonable request to the corresponding author.



\bibliographystyle{mnras}
\bibliography{paper} 

\begin{thebibliography}{}
\makeatletter
\relax
\def\mn@urlcharsother{\let\do\@makeother \do\$\do\&\do\#\do\^\do\_\do\%\do\~}
\def\mn@doi{\begingroup\mn@urlcharsother \@ifnextchar [ {\mn@doi@}
  {\mn@doi@[]}}
\def\mn@doi@[#1]#2{\def\@tempa{#1}\ifx\@tempa\@empty \href
  {http://dx.doi.org/#2} {doi:#2}\else \href {http://dx.doi.org/#2} {#1}\fi
  \endgroup}
\def\mn@eprint#1#2{\mn@eprint@#1:#2::\@nil}
\def\mn@eprint@arXiv#1{\href {http://arxiv.org/abs/#1} {{\tt arXiv:#1}}}
\def\mn@eprint@dblp#1{\href {http://dblp.uni-trier.de/rec/bibtex/#1.xml}
  {dblp:#1}}
\def\mn@eprint@#1:#2:#3:#4\@nil{\def\@tempa {#1}\def\@tempb {#2}\def\@tempc
  {#3}\ifx \@tempc \@empty \let \@tempc \@tempb \let \@tempb \@tempa \fi \ifx
  \@tempb \@empty \def\@tempb {arXiv}\fi \@ifundefined
  {mn@eprint@\@tempb}{\@tempb:\@tempc}{\expandafter \expandafter \csname
  mn@eprint@\@tempb\endcsname \expandafter{\@tempc}}}

\bibitem[\protect\citeauthoryear{{Ataiee}, {Baruteau}, {Alibert}  \&
  {Benz}}{{Ataiee} et~al.}{2018}]{Ataiee_etal2018}
{Ataiee} S.,  {Baruteau} C.,  {Alibert} Y.,   {Benz} W.,  2018, \mn@doi [\aap]
  {10.1051/0004-6361/201732026}, \href
  {https://ui.adsabs.harvard.edu/abs/2018A&A...615A.110A} {615, A110}

\bibitem[\protect\citeauthoryear{{Baruteau} \& {Zhu}}{{Baruteau} \&
  {Zhu}}{2016}]{BZ2016}
{Baruteau} C.,  {Zhu} Z.,  2016, \mn@doi [\mnras] {10.1093/mnras/stv2527},
  \href {https://ui.adsabs.harvard.edu/abs/2016MNRAS.458.3927B} {458, 3927}

\bibitem[\protect\citeauthoryear{{Baruteau}, {Bai}, {Mordasini}  \&
  {Molli{\`e}re}}{{Baruteau} et~al.}{2016}]{2016SSRv..205...77B}
{Baruteau} C.,  {Bai} X.,  {Mordasini} C.,   {Molli{\`e}re} P.,  2016, \mn@doi
  [\ssr] {10.1007/s11214-016-0258-z}, \href
  {https://ui.adsabs.harvard.edu/abs/2016SSRv..205...77B} {205, 77}

\bibitem[\protect\citeauthoryear{{Baruteau}, {Wafflard-Fernandez}, {Le Gal},
  {Debras}, {Carmona}, {Fuente}  \& {Rivi{\`e}re-Marichalar}}{{Baruteau}
  et~al.}{2021}]{Baruteau2021}
{Baruteau} C.,  {Wafflard-Fernandez} G.,  {Le Gal} R.,  {Debras} F.,  {Carmona}
  A.,  {Fuente} A.,   {Rivi{\`e}re-Marichalar} P.,  2021, \mn@doi [\mnras]
  {10.1093/mnras/stab1045}, \href
  {https://ui-adsabs-harvard-edu.insu.bib.cnrs.fr/abs/2021MNRAS.505..359B}
  {505, 359}

\bibitem[\protect\citeauthoryear{{Bitsch}}{{Bitsch}}{2019}]{2019A&A...630A..51B}
{Bitsch} B.,  2019, \mn@doi [\aap] {10.1051/0004-6361/201935877}, \href
  {https://ui.adsabs.harvard.edu/abs/2019A&A...630A..51B} {630, A51}

\bibitem[\protect\citeauthoryear{{Bitsch} \& {Kley}}{{Bitsch} \&
  {Kley}}{2010}]{2010A&A...523A..30B}
{Bitsch} B.,  {Kley} W.,  2010, \mn@doi [\aap] {10.1051/0004-6361/201014414},
  \href {https://ui.adsabs.harvard.edu/abs/2010A&A...523A..30B} {523, A30}

\bibitem[\protect\citeauthoryear{{Bitsch}, {Crida}, {Libert}  \&
  {Lega}}{{Bitsch} et~al.}{2013}]{2013A&A...555A.124B}
{Bitsch} B.,  {Crida} A.,  {Libert} A.~S.,   {Lega} E.,  2013, \mn@doi [\aap]
  {10.1051/0004-6361/201220310}, \href
  {https://ui.adsabs.harvard.edu/abs/2013A&A...555A.124B} {555, A124}

\bibitem[\protect\citeauthoryear{{Bitsch}, {Morbidelli}, {Johansen}, {Lega},
  {Lambrechts}  \& {Crida}}{{Bitsch} et~al.}{2018}]{Bitsch_etal2018}
{Bitsch} B.,  {Morbidelli} A.,  {Johansen} A.,  {Lega} E.,  {Lambrechts} M.,
  {Crida} A.,  2018, \mn@doi [\aap] {10.1051/0004-6361/201731931}, \href
  {https://ui.adsabs.harvard.edu/abs/2018A&A...612A..30B} {612, A30}

\bibitem[\protect\citeauthoryear{{Bitsch}, {Izidoro}, {Johansen}, {Raymond},
  {Morbidelli}, {Lambrechts}  \& {Jacobson}}{{Bitsch}
  et~al.}{2019}]{2019A&A...623A..88B}
{Bitsch} B.,  {Izidoro} A.,  {Johansen} A.,  {Raymond} S.~N.,  {Morbidelli} A.,
   {Lambrechts} M.,   {Jacobson} S.~A.,  2019, \mn@doi [\aap]
  {10.1051/0004-6361/201834489}, \href
  {https://ui.adsabs.harvard.edu/abs/2019A&A...623A..88B} {623, A88}

\bibitem[\protect\citeauthoryear{{Bitsch}, {Trifonov}  \& {Izidoro}}{{Bitsch}
  et~al.}{2020}]{2020A&A...643A..66B}
{Bitsch} B.,  {Trifonov} T.,   {Izidoro} A.,  2020, \mn@doi [\aap]
  {10.1051/0004-6361/202038856}, \href
  {https://ui.adsabs.harvard.edu/abs/2020A&A...643A..66B} {643, A66}

\bibitem[\protect\citeauthoryear{{Brasser}, {Bitsch}  \& {Matsumura}}{{Brasser}
  et~al.}{2017}]{2017AJ....153..222B}
{Brasser} R.,  {Bitsch} B.,   {Matsumura} S.,  2017, \mn@doi [\aj]
  {10.3847/1538-3881/aa6ba3}, \href
  {https://ui.adsabs.harvard.edu/abs/2017AJ....153..222B} {153, 222}

\bibitem[\protect\citeauthoryear{{Chambers}}{{Chambers}}{2016}]{2016ApJ...825...63C}
{Chambers} J.~E.,  2016, \mn@doi [\apj] {10.3847/0004-637X/825/1/63}, \href
  {https://ui.adsabs.harvard.edu/abs/2016ApJ...825...63C} {825, 63}

\bibitem[\protect\citeauthoryear{{Chen}, {Wang}, {Li}, {Baruteau}  \&
  {Lin}}{{Chen} et~al.}{2021}]{YXChen21}
{Chen} Y.-X.,  {Wang} Z.,  {Li} Y.-P.,  {Baruteau} C.,   {Lin} D. N.~C.,  2021,
  \mn@doi [\apj] {10.3847/1538-4357/ac23d7}, \href
  {https://ui-adsabs-harvard-edu.insu.bib.cnrs.fr/abs/2021ApJ...922..184C}
  {922, 184}

\bibitem[\protect\citeauthoryear{{Chrenko}, {Bro{\v{z}}}  \&
  {Lambrechts}}{{Chrenko} et~al.}{2017}]{Chrenko2017}
{Chrenko} O.,  {Bro{\v{z}}} M.,   {Lambrechts} M.,  2017, \mn@doi [\aap]
  {10.1051/0004-6361/201731033}, \href
  {https://ui-adsabs-harvard-edu.insu.bib.cnrs.fr/abs/2017A&A...606A.114C}
  {606, A114}

\bibitem[\protect\citeauthoryear{{Cresswell} \& {Nelson}}{{Cresswell} \&
  {Nelson}}{2008}]{2008A&A...482..677C}
{Cresswell} P.,  {Nelson} R.~P.,  2008, \mn@doi [\aap]
  {10.1051/0004-6361:20079178}, \href
  {https://ui.adsabs.harvard.edu/abs/2008A&A...482..677C} {482, 677}

\bibitem[\protect\citeauthoryear{{Crida}, {Morbidelli}  \& {Masset}}{{Crida}
  et~al.}{2006}]{Crida_etal2006}
{Crida} A.,  {Morbidelli} A.,   {Masset} F.,  2006, \mn@doi [\icarus]
  {10.1016/j.icarus.2005.10.007}, \href
  {https://ui.adsabs.harvard.edu/abs/2006Icar..181..587C} {181, 587}

\bibitem[\protect\citeauthoryear{{D'Angelo}, {Lubow}  \& {Bate}}{{D'Angelo}
  et~al.}{2006}]{2006ApJ...652.1698D}
{D'Angelo} G.,  {Lubow} S.~H.,   {Bate} M.~R.,  2006, \mn@doi [\apj]
  {10.1086/508451}, \href
  {https://ui.adsabs.harvard.edu/abs/2006ApJ...652.1698D} {652, 1698}

\bibitem[\protect\citeauthoryear{{Debras}, {Baruteau}  \& {Donati}}{{Debras}
  et~al.}{2021}]{Debras21}
{Debras} F.,  {Baruteau} C.,   {Donati} J.-F.,  2021, \mn@doi [\mnras]
  {10.1093/mnras/staa3397}, \href
  {https://ui-adsabs-harvard-edu.insu.bib.cnrs.fr/abs/2021MNRAS.500.1621D}
  {500, 1621}

\bibitem[\protect\citeauthoryear{{Duffell}}{{Duffell}}{2015}]{2015ApJ...807L..11D}
{Duffell} P.~C.,  2015, \mn@doi [\apjl] {10.1088/2041-8205/807/1/L11}, \href
  {https://ui.adsabs.harvard.edu/abs/2015ApJ...807L..11D} {807, L11}

\bibitem[\protect\citeauthoryear{{Eklund} \& {Masset}}{{Eklund} \&
  {Masset}}{2017}]{2017MNRAS.469..206E}
{Eklund} H.,  {Masset} F.~S.,  2017, \mn@doi [\mnras] {10.1093/mnras/stx856},
  \href {https://ui.adsabs.harvard.edu/abs/2017MNRAS.469..206E} {469, 206}

\bibitem[\protect\citeauthoryear{{Fang} \& {Margot}}{{Fang} \&
  {Margot}}{2012}]{2012ApJ...761...92F}
{Fang} J.,  {Margot} J.-L.,  2012, \mn@doi [\apj] {10.1088/0004-637X/761/2/92},
  \href {https://ui.adsabs.harvard.edu/abs/2012ApJ...761...92F} {761, 92}

\bibitem[\protect\citeauthoryear{{Fromenteau} \& {Masset}}{{Fromenteau} \&
  {Masset}}{2019}]{2019MNRAS.485.5035F}
{Fromenteau} S.,  {Masset} F.~S.,  2019, \mn@doi [\mnras]
  {10.1093/mnras/stz718}, \href
  {https://ui.adsabs.harvard.edu/abs/2019MNRAS.485.5035F} {485, 5035}

\bibitem[\protect\citeauthoryear{{Izidoro}, {Bitsch}, {Raymond}, {Johansen},
  {Morbidelli}, {Lambrechts}  \& {Jacobson}}{{Izidoro}
  et~al.}{2021}]{Izidoro21}
{Izidoro} A.,  {Bitsch} B.,  {Raymond} S.~N.,  {Johansen} A.,  {Morbidelli} A.,
   {Lambrechts} M.,   {Jacobson} S.~A.,  2021, \mn@doi [\aap]
  {10.1051/0004-6361/201935336}, \href
  {https://ui.adsabs.harvard.edu/abs/2021A&A...650A.152I} {650, A152}

\bibitem[\protect\citeauthoryear{{Johansen} \& {Lambrechts}}{{Johansen} \&
  {Lambrechts}}{2017}]{2017AREPS..45..359J}
{Johansen} A.,  {Lambrechts} M.,  2017, \mn@doi [Annual Review of Earth and
  Planetary Sciences] {10.1146/annurev-earth-063016-020226}, \href
  {https://ui.adsabs.harvard.edu/abs/2017AREPS..45..359J} {45, 359}

\bibitem[\protect\citeauthoryear{{Johansen}, {Davies}, {Church}  \&
  {Holmelin}}{{Johansen} et~al.}{2012}]{2012ApJ...758...39J}
{Johansen} A.,  {Davies} M.~B.,  {Church} R.~P.,   {Holmelin} V.,  2012,
  \mn@doi [\apj] {10.1088/0004-637X/758/1/39}, \href
  {https://ui.adsabs.harvard.edu/abs/2012ApJ...758...39J} {758, 39}

\bibitem[\protect\citeauthoryear{{Kley} \& {Dirksen}}{{Kley} \&
  {Dirksen}}{2006}]{2006A&A...447..369K}
{Kley} W.,  {Dirksen} G.,  2006, \mn@doi [\aap] {10.1051/0004-6361:20053914},
  \href {https://ui.adsabs.harvard.edu/abs/2006A&A...447..369K} {447, 369}

\bibitem[\protect\citeauthoryear{{Kley}, {Peitz}  \& {Bryden}}{{Kley}
  et~al.}{2004}]{Kley04}
{Kley} W.,  {Peitz} J.,   {Bryden} G.,  2004, \mn@doi [\aap]
  {10.1051/0004-6361:20031589}, \href
  {https://ui-adsabs-harvard-edu.insu.bib.cnrs.fr/abs/2004A&A...414..735K}
  {414, 735}

\bibitem[\protect\citeauthoryear{{Kruijer}, {Burkhardt}, {Budde}  \&
  {Kleine}}{{Kruijer} et~al.}{2017}]{Kruijer2017}
{Kruijer} T.~S.,  {Burkhardt} C.,  {Budde} G.,   {Kleine} T.,  2017, \mn@doi
  [Proceedings of the National Academy of Science] {10.1073/pnas.1704461114},
  \href {https://ui.adsabs.harvard.edu/abs/2017PNAS..114.6712K} {114, 6712}

\bibitem[\protect\citeauthoryear{{Lambrechts} \& {Johansen}}{{Lambrechts} \&
  {Johansen}}{2014}]{2014A&A...572A.107L}
{Lambrechts} M.,  {Johansen} A.,  2014, \mn@doi [\aap]
  {10.1051/0004-6361/201424343}, \href
  {https://ui.adsabs.harvard.edu/abs/2014A&A...572A.107L} {572, A107}

\bibitem[\protect\citeauthoryear{{Lambrechts}, {Johansen}  \&
  {Morbidelli}}{{Lambrechts} et~al.}{2014}]{Lambrechts_etal2014}
{Lambrechts} M.,  {Johansen} A.,   {Morbidelli} A.,  2014, \mn@doi [\aap]
  {10.1051/0004-6361/201423814}, \href
  {https://ui.adsabs.harvard.edu/abs/2014A&A...572A..35L} {572, A35}

\bibitem[\protect\citeauthoryear{{Lambrechts}, {Morbidelli}, {Jacobson},
  {Johansen}, {Bitsch}, {Izidoro}  \& {Raymond}}{{Lambrechts}
  et~al.}{2019}]{2019A&A...627A..83L}
{Lambrechts} M.,  {Morbidelli} A.,  {Jacobson} S.~A.,  {Johansen} A.,  {Bitsch}
  B.,  {Izidoro} A.,   {Raymond} S.~N.,  2019, \mn@doi [\aap]
  {10.1051/0004-6361/201834229}, \href
  {https://ui.adsabs.harvard.edu/abs/2019A&A...627A..83L} {627, A83}

\bibitem[\protect\citeauthoryear{Levison, Kretke  \& Duncan}{Levison
  et~al.}{2015}]{PMID:26289203}
Levison H.~F.,  Kretke K.~A.,   Duncan M.~J.,  2015, \mn@doi [Nature]
  {10.1038/nature14675}, 524, 322

\bibitem[\protect\citeauthoryear{{Lissauer} et~al.,}{{Lissauer}
  et~al.}{2011}]{2011ApJS..197....8L}
{Lissauer} J.~J.,  et~al., 2011, \mn@doi [\apjs] {10.1088/0067-0049/197/1/8},
  \href {https://ui.adsabs.harvard.edu/abs/2011ApJS..197....8L} {197, 8}

\bibitem[\protect\citeauthoryear{{Liu} \& {Ormel}}{{Liu} \&
  {Ormel}}{2018}]{LO2018}
{Liu} B.,  {Ormel} C.~W.,  2018, \mn@doi [\aap] {10.1051/0004-6361/201732307},
  \href {https://ui.adsabs.harvard.edu/abs/2018A&A...615A.138L} {615, A138}

\bibitem[\protect\citeauthoryear{{Masset}}{{Masset}}{2000}]{Masset2000}
{Masset} F.,  2000, \mn@doi [\aaps] {10.1051/aas:2000116}, \href
  {https://ui.adsabs.harvard.edu/abs/2000A&AS..141..165M} {141, 165}

\bibitem[\protect\citeauthoryear{{Matsumura}, {Brasser}  \& {Ida}}{{Matsumura}
  et~al.}{2021}]{2021A&A...650A.116M}
{Matsumura} S.,  {Brasser} R.,   {Ida} S.,  2021, \mn@doi [\aap]
  {10.1051/0004-6361/202039210}, \href
  {https://ui.adsabs.harvard.edu/abs/2021A&A...650A.116M} {650, A116}

\bibitem[\protect\citeauthoryear{{Mayor} et~al.,}{{Mayor}
  et~al.}{2011}]{2011arXiv1109.2497M}
{Mayor} M.,  et~al., 2011, arXiv e-prints, \href
  {https://ui.adsabs.harvard.edu/abs/2011arXiv1109.2497M} {p. arXiv:1109.2497}

\bibitem[\protect\citeauthoryear{{Morbidelli} \& {Nesvorny}}{{Morbidelli} \&
  {Nesvorny}}{2012}]{2012A&A...546A..18M}
{Morbidelli} A.,  {Nesvorny} D.,  2012, \mn@doi [\aap]
  {10.1051/0004-6361/201219824}, \href
  {https://ui.adsabs.harvard.edu/abs/2012A&A...546A..18M} {546, A18}

\bibitem[\protect\citeauthoryear{{Ormel}, {Vazan}  \& {Brouwers}}{{Ormel}
  et~al.}{2021}]{2021A&A...647A.175O}
{Ormel} C.~W.,  {Vazan} A.,   {Brouwers} M.~G.,  2021, \mn@doi [\aap]
  {10.1051/0004-6361/202039706}, \href
  {https://ui.adsabs.harvard.edu/abs/2021A&A...647A.175O} {647, A175}

\bibitem[\protect\citeauthoryear{{Paardekooper} \& {Mellema}}{{Paardekooper} \&
  {Mellema}}{2006}]{2006A&A...459L..17P}
{Paardekooper} S.~J.,  {Mellema} G.,  2006, \mn@doi [\aap]
  {10.1051/0004-6361:20066304}, \href
  {https://ui.adsabs.harvard.edu/abs/2006A&A...459L..17P} {459, L17}

\bibitem[\protect\citeauthoryear{{Papaloizou}, {Nelson}  \&
  {Masset}}{{Papaloizou} et~al.}{2001}]{2001A&A...366..263P}
{Papaloizou} J.~C.~B.,  {Nelson} R.~P.,   {Masset} F.,  2001, \mn@doi [\aap]
  {10.1051/0004-6361:20000011}, \href
  {https://ui.adsabs.harvard.edu/abs/2001A&A...366..263P} {366, 263}

\bibitem[\protect\citeauthoryear{{Pierens} \& {Nelson}}{{Pierens} \&
  {Nelson}}{2008}]{Pierens08}
{Pierens} A.,  {Nelson} R.~P.,  2008, \mn@doi [\aap]
  {10.1051/0004-6361:20079062}, \href
  {https://ui-adsabs-harvard-edu.insu.bib.cnrs.fr/abs/2008A&A...482..333P}
  {482, 333}

\bibitem[\protect\citeauthoryear{{Romero}, {Masset}  \& {Teyssier}}{{Romero}
  et~al.}{2021}]{2021MNRAS.tmp.3059R}
{Romero} D. A.~V.,  {Masset} F.~S.,   {Teyssier} R.,  2021, \mn@doi [\mnras]
  {10.1093/mnras/stab3334}, \href
  {https://ui.adsabs.harvard.edu/abs/2021MNRAS.tmp.3059R} {}

\bibitem[\protect\citeauthoryear{{Wu}}{{Wu}}{2019}]{2019ApJ...874...91W}
{Wu} Y.,  2019, \mn@doi [\apj] {10.3847/1538-4357/ab06f8}, \href
  {https://ui.adsabs.harvard.edu/abs/2019ApJ...874...91W} {874, 91}

\bibitem[\protect\citeauthoryear{{Xie} et~al.,}{{Xie}
  et~al.}{2016}]{2016PNAS..11311431X}
{Xie} J.-W.,  et~al., 2016, \mn@doi [Proceedings of the National Academy of
  Science] {10.1073/pnas.1604692113}, \href
  {https://ui.adsabs.harvard.edu/abs/2016PNAS..11311431X} {113, 11431}

\bibitem[\protect\citeauthoryear{{Zhang} et~al.,}{{Zhang}
  et~al.}{2018}]{Zhang2018}
{Zhang} S.,  et~al., 2018, \mn@doi [\apjl] {10.3847/2041-8213/aaf744}, \href
  {https://ui.adsabs.harvard.edu/abs/2018ApJ...869L..47Z} {869, L47}

\bibitem[\protect\citeauthoryear{{Zhu}, {Petrovich}, {Wu}, {Dong}  \&
  {Xie}}{{Zhu} et~al.}{2018}]{2018ApJ...860..101Z}
{Zhu} W.,  {Petrovich} C.,  {Wu} Y.,  {Dong} S.,   {Xie} J.,  2018, \mn@doi
  [\apj] {10.3847/1538-4357/aac6d5}, \href
  {https://ui.adsabs.harvard.edu/abs/2018ApJ...860..101Z} {860, 101}

\bibitem[\protect\citeauthoryear{{de Val-Borro} et~al.,}{{de Val-Borro}
  et~al.}{2006}]{deVal_etal2006}
{de Val-Borro} M.,  et~al., 2006, \mn@doi [\mnras]
  {10.1111/j.1365-2966.2006.10488.x}, \href
  {https://ui.adsabs.harvard.edu/abs/2006MNRAS.370..529D} {370, 529}

\makeatother
\end{thebibliography}






\bsp	
\label{lastpage}
\end{document}